\documentclass[aip,jcp,preprint]{revtex4-1}

\usepackage[T1]{fontenc}       

\usepackage{graphicx}
\usepackage{dcolumn}
\usepackage{bm}
\usepackage{amsmath}
\usepackage{amsfonts}
\usepackage{amssymb}
\usepackage{braket}
\usepackage[separate-uncertainty = true,multi-part-units=single]{siunitx}
\usepackage{hyperref}
\usepackage{subcaption}
\usepackage{color}

\let\oldsqrt\sqrt
\def\sqrt{\mathpalette\DHLhksqrt}
\def\DHLhksqrt#1#2{%
\setbox0=\hbox{$#1\oldsqrt{#2\,}$}\dimen0=\ht0
\advance\dimen0-0.2\ht0
\setbox2=\hbox{\vrule height\ht0 depth -\dimen0}%
{\box0\lower0.4pt\box2}}

\begin{document}
\preprint{adts.201800078}
\title{5D Entanglement in Star Polymer Dynamics} 

\author{Airidas Korolkovas}
\affiliation{Institut Laue-Langevin, 71 rue des Martyrs, 38000 Grenoble, France}
\affiliation{Department of Physics and Astronomy, Uppsala University, Lägerhyddsvägen 1, 752 37 Uppsala, Sweden}
\email{korolkovas@ill.fr}

\date{\today}

\begin{abstract}
Star polymers are within the most topologically entangled macromolecules. For a star to move the current theory is that one arm must retract to the branch point. The probability of this event falls exponentially with molecular weight, and a quicker relaxation pathway eventually takes over. With a simulation over a hundred times faster than earlier studies, we demonstrate that the mean square displacement scales with a power law 1/16 in time, instead of the previously assumed zero. It suggests that star polymer motion is the result of two linear relaxations coinciding in time. By analogy to linear polymers, which reptate with a random walk embedded in a 3D network, we show that star polymers relax by a random walk in a 5D network.
\end{abstract}

\maketitle

\section{Introduction}
\begin{figure}[ptbh!] 
\includegraphics[width=0.6\linewidth]{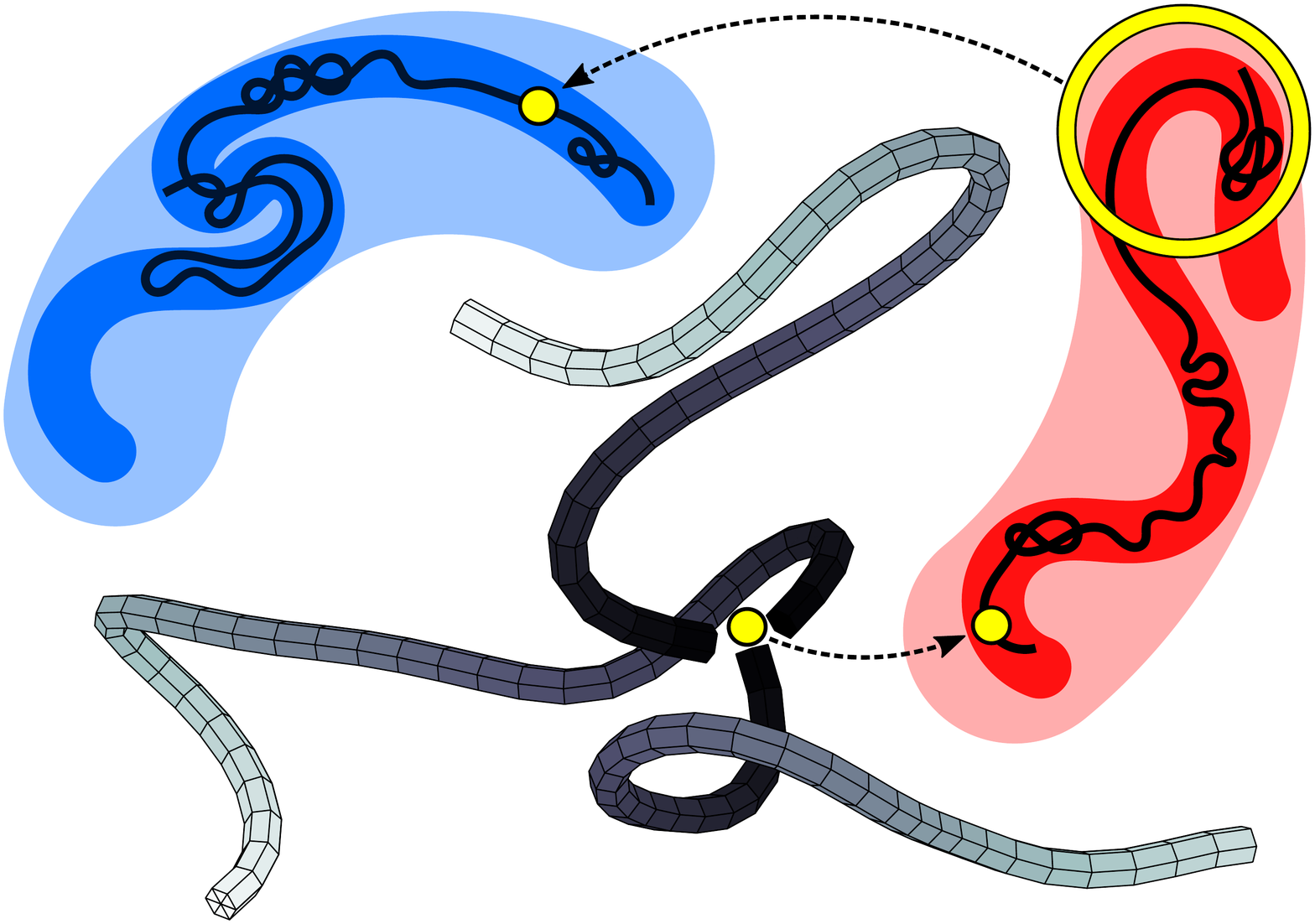}
\caption{The branch point of an entangled star polymer moves whenever two reptation attempts of its linear arms coincide. The result is a random walk, nested in five other random walks.}\label{5d}
\end{figure}

Entangled polymers are a unique form of matter that continues to puzzle physicists. On the small scale of a monomer the motion is very fast and is similar to simple, or Newtonian liquids. On the large scale of the whole chain, the effective diffusion is extremely slow, accompanied by a huge viscosity and the ability to store elastic energy like a rubber band. It is rooted in the excluded volume \emph{repulsion} between the long chains, preventing them to cross each other's path. This topological effect, called entanglement, has proven difficult to treat theoretically. Its nature is fundamentally different from many other thick fluids, like bitumen~\cite{edgeworth1984pitch}, whose viscosity derives from strong \emph{attractive} forces of enthalpic origin. Polymers, in contrast, are mostly governed by their configurational entropy. It is the reason why rubber (a cross-linked polymer network) expands upon cooling, and contracts upon heating, the opposite of conventional expectations.

The central goal of polymer dynamics is to bridge the understanding of molecular motion between the small and the large scales. For linear polymers, the most accepted description so far is the tube theory, which focuses on a single chain, confined by the mean field of all other chains, in a self-consistent way. With some modifications, called GLaMM, the theory can fit rheological experiments~\cite{graham2003microscopic}, although the structural response far from equilibrium is still debated~\cite{wang2017fingerprinting}. An essential ingredient of tube theory is called reptation, confirmed by neutron spin-echo (NSE) spectroscopy~\cite{richter2005neutron}  and computer simulations~\cite{kremer1990dynamics}. Its main signature is the mean square displacement scaling with a power law 1/4 in time. It results from the monomer moving with a random walk, inside a random walking chain, inside a tube shaped like a random walk. In other words, it is a one dimensional diffusion, embedded in a three dimensional confining network~\cite{de1971reptation}. The topological aspect of entangled polymers bears fundamental similarities to the quantum Hall effect~\cite{kholodenko1994tube}.

Branched polymers, and stars in particular~\cite{grest1996star}, are becoming more accessible, thanks to recent innovations in their chemical synthesis~\cite{hadjichristidis2001polymers, ren2016star}. They are already used as additives in consumer products (viscosity modifiers, defoaming agents) and in biomedical research as drug delivery vehicles. While the static structure of star polymer systems has been well studied in both bulk~\cite{likos1998star} and interfaces~\cite{taddese2015thermodynamics}, their dynamics pose considerable challenges, as they can be many orders of magnitude slower compared to linear chains of the same molecular weight $N$. It has been recognized that reptation is impossible for branched polymers, and a new relaxation pathway of arm retraction has been proposed~\cite{klein1986dynamics, milner1997parameter, mcleish2002tube}. Its main outcome is the exponential growth of the relaxation time $\tau_a \propto e^{N/N_a}$, where $N_a$ is the arm length at the onset of star dynamics. The exponential scaling was first observed in rheology and diffusion experiments~\cite{bartels1986self}. At the microscopic scale of the monomer, star dynamics have been probed by NSE~\cite{holler2017role} and computer simulations~\cite{bacova2013dynamics}. The current consensus is in favor of arm retraction, although a broad mesoscopic time range remains unexplored and some open questions remain~\cite{cao2016microscopic, zhu2017arm}.

In this work we present a multi-chain simulation running over a 100 times faster than the best supercomputer result of a comparable star polymer problem published to date~\cite{bacova2013dynamics}. The speedup quoted here is an order-of-magnitude estimate, whose precise value cannot be guaranteed, as it depends on the specific hardware, driver, compiler, and kernel optimization, the random number generator, as well as details of the physical model such as the algorithm used to constrain the bond length and (optionally) the bond stiffness. A large part of the speedup, about $\times 10$, is due to a longer time step, made possible by coarse-graining and simplifying the polymer physics to the bare minimum, just enough to suppress random chain crossings. This model was presented in our earlier simulation on linear chains at equilibrium~\cite{korolkovas2016simulation} and under shear flow~\cite{korolkovas2018dynamical}. The second part of the speedup, roughly the remaining $\times 10$, is due to our exploitation of a unique polymer property: beads along a chain are always close to each other. Thanks to their intrinsically lower entropy, polymers require only a fraction of the computing effort needed for small molecules. To enable this idea in practice, our novelty is to take full advantage of texture mapping, a special kind of memory used in graphics processing units (GPU). While some amount of GPU acceleration is available in most molecular dynamics software (LAAMPS~\cite{glaser2015strong}, Espresso++~\cite{arnold2013espresso}, GROMACS~\cite{abraham2015gromacs}, Amber~\cite{salomon2013routine}), so far none of them have tapped into the potential of textures. Their original usage is to generate 3D graphics, and here we  have repurposed them to map continuous polymers in and out of a discrete grid.

Our code can be adapted to many situations involving coarse-grained polymers (i.e. interfacial brush flow~\cite{korolkovas2017polymer}), and its most striking advantage is for small systems, up to $10^5$ particles, that fit into a single GPU cycle. Entangled star polymers have ludicrously long relaxation times already at that size. Probing new physics in this case requires higher speed, not a bigger box, and that is where our code excels. Other situations like linear or ring polymers have much faster relaxation times, so they would need a multi-GPU cluster to access beyond the currently known physics, and we leave that for a future study.

For a three-arm star polymer melt, we reproduce existing data and shed ample new light on the time scales beyond. We find that for sufficiently big stars, the mean square displacement of the branch point scales with $t^{1/16}$ in time. In addition, the arms are universally observed to relax with $t^{-2/3}$ in time. These new power laws are not accounted for in existing polymer theories, inviting further developments. We propose a new relaxation pathway, depicted in Fig.~\ref{5d}. Each arm is attempting to reptate at the rate of $t^{1/4}$, and two such reptations will coincide by chance at the rate of ${\left(t^{1/4}\right)}^{1/4} = t^{1/16}$. The overall motion is made up of six random walks nested in each other, or equivalently, a one dimensional random walk confined by a five dimensional network of topological entanglement. For very big stars, this mode of relaxation dominates over the much slower exponential rate predicted by arm retraction theory. There exists a relationship between linear polymer reptation and quantum chaotic systems~\cite{kholodenko1996reptation}, and it remains to be seen if the analogy can be extended for branched polymers.

\section{Computational Section}
\begin{figure}[ptbh!] 
\includegraphics[width=0.8\linewidth]{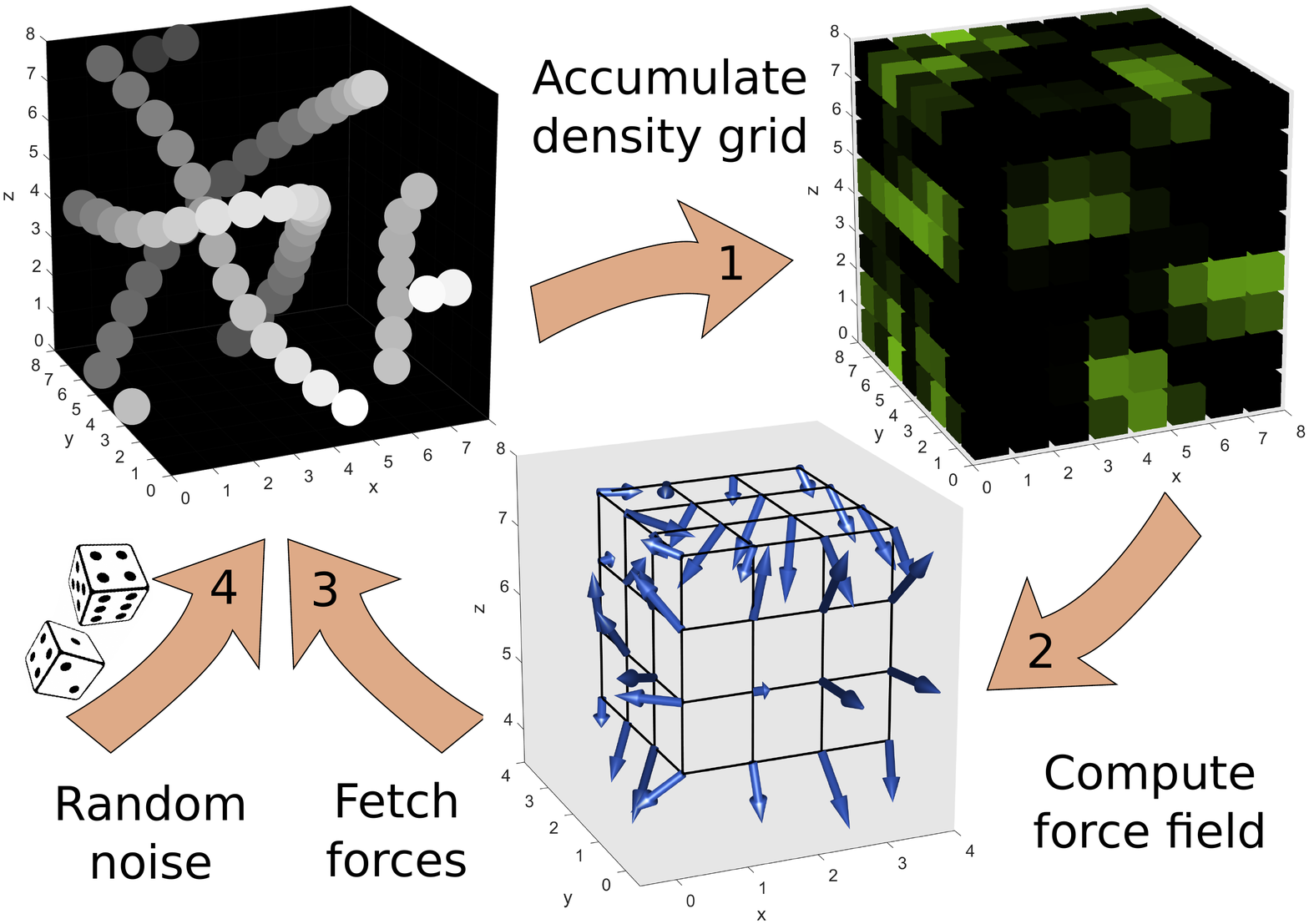}
\caption{Simulation algorithm. Polymer coordinates are stored in continuous space (top left). 1) A density field is constructed on the grid (top right). 2) The density is convoluted with a Gaussian potential to produce the force field (bottom). 3) The forces are fetched from the field. 4) Thermal noise is added. 5) The bond length is constrained to less than one bin wide (not shown). The speed is 5000 steps/s with 64000 beads.}\label{algorithm}
\end{figure}

The slowest part of the simulation is the non-bonded, or the excluded volume force calculation. Our objective is to boost it to the edge of what is possible with today's technology. The algorithm is illustrated in Fig.~\ref{algorithm}. It is based on the particle-in-cell (PIC) method. Instead of calculating every single interaction as in molecular dynamics (MD), the particles are binned to a grid, from which the net force is read off. This scheme gains speed at the expense of lower spatial resolution. The small loss of accuracy is generally acceptable for long range forces in gravity and plasma simulations~\cite{hariri2016portable}, where this technique was originally devised. More recently, PIC has gained popularity for coarse-grained simulation of block copolymers~\cite{glaser2014universality} and soft matter in general~\cite{zhang2013new}. It has sufficient accuracy for polymeric liquids above the glass transition temperature $T_g$. We do not consider cases below $T_g$, as that would require the resolution of hardcore atomic forces. The performance of PIC is ultimately capped by the latency of data transfer from continuous space to the grid and back. Typically, merely 1\% of the peak GPU memory bandwidth can be utilized, due to disorganized locations of small particles. Long polymers, on the other hand, are composed of spatially correlated beads, making them an ideal candidate for texture memory. This special hardware can eschew the regular memory bottleneck in cases when consecutive threads operate on nearby bins. We use texture fetching (Fig.~\ref{algorithm}-3) to return all 3 force components, with linear interpolation from the 8 nearest bins, in just one instruction. Writing into textures with interpolation efficiently (Fig.~\ref{algorithm}-1) was achieved with a two-step procedure involving some technicalities, explained in our source code available for download from Zenodo Ref.~\cite{zenodoref}.

\begin{figure}[tbh] 
\includegraphics[width=0.5\linewidth]{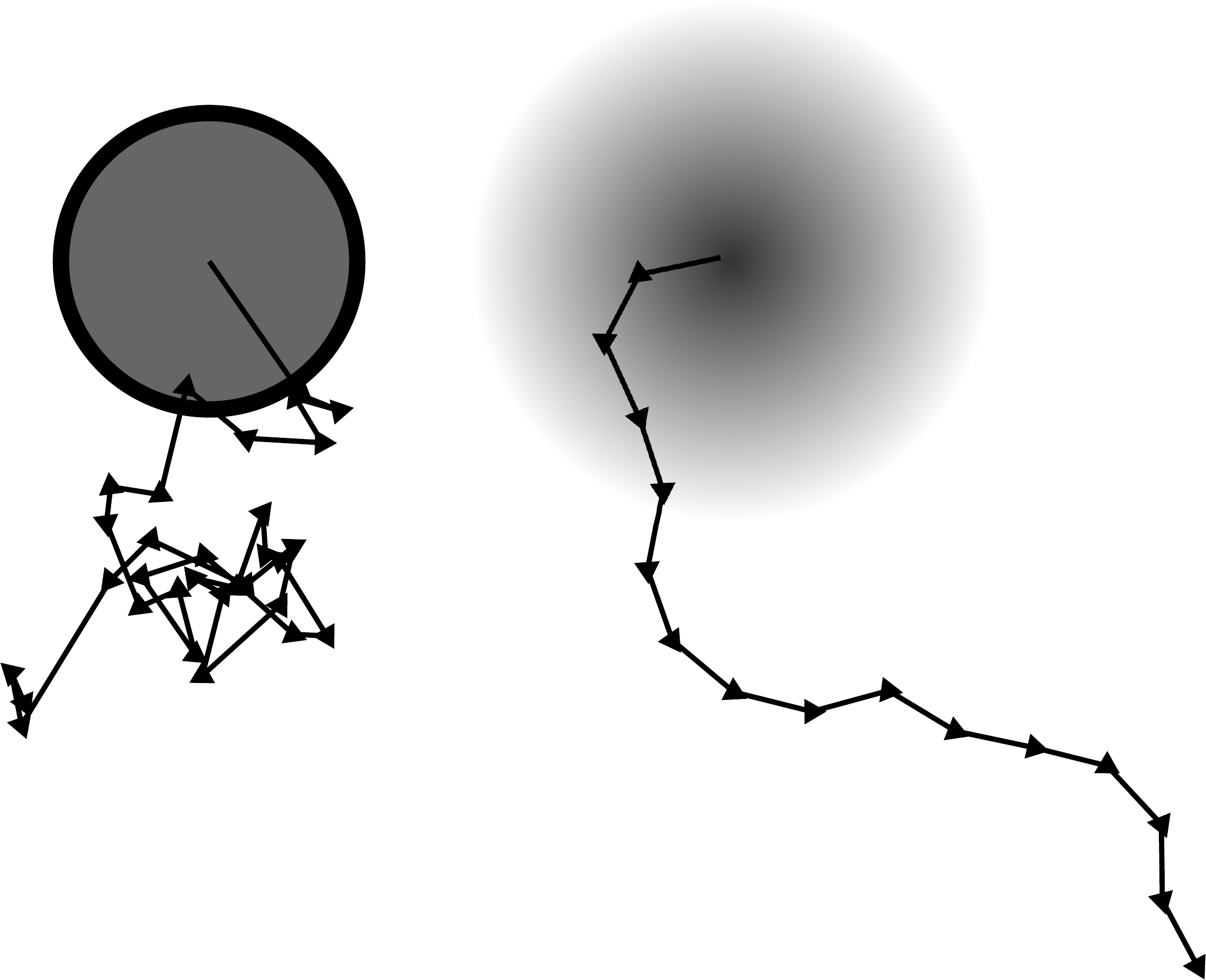}
\caption{Discrete random walk samples: uncorrelated for a hard particle (left), gently decorrelating for a soft particle (right).}\label{rwalkers}
\end{figure}

The force on a bin is a weighted sum of the density $\rho$ around it (Fig.~\ref{algorithm}-2). This step, called convolution, is very fast when using only the nearest $3\times 3 \times 3$ bins. The weights are chosen to match the Gaussian potential function:
\begin{subequations}
\begin{align}\label{Gaussian}
\mathbf{F} &= -\nabla U(\mathbf{r}) = \frac{k_B T \mathbf{r}}{\lambda^2}\exp \left(-\frac{\mathbf{r}^2}{2\lambda^2} \right)\\
& \approx \frac{k_B T}{\lambda^2} \sum_{i,j,k=-1}^1 \rho_{ijk}w_{ijk} (i\hat{\mathbf{x}} + j\hat{\mathbf{y}} + k\hat{\mathbf{z}})
\end{align}
\end{subequations}
The bin size is set equal to $\lambda$, so the three distinct weights are $w_{100} = e^{-1/2}$, $w_{110} = e^{-1}$, and $w_{111} = e^{-3/2}$. The bead position $\mathbf{R}$ moves in time $t$ according to the Brownian equation of motion:
\begin{equation}
\zeta \frac{d\mathbf{R}}{dt} = \mathbf{F} + \sqrt{6\zeta k_B T} \mathbf{W}(t)
\end{equation}
where $\zeta$ is the effective friction coefficient, and $\mathbf{W}$ is a unit vector of random orientation. Mathematically, the random force is assumed to fluctuate at an infinite frequency: $\braket{\mathbf{W}(t)\cdot \mathbf{W}(t')} = \delta(t-t')$ (the Wiener process). In computer simulations, it must be truncated to a finite duration, usually equal to one time step $\braket{\mathbf{W}(t)\cdot \mathbf{W}(t')} = \delta_{t,t'}/\Delta t$ (the Markov chain). This model is used to describe hard, atomistic particles, depicted in Fig.~\ref{rwalkers} (left). It cannot be applied to our soft blobs, since the strength of such a random force $\sqrt{6\zeta k_B T/\Delta t} \approx \lambda$ is similar to that of the Gaussian barrier, Eq.~\eqref{Gaussian}. This would cause unphysical chain crossings and sabotage the entanglement dynamics. Our solution is to use a gently decorrelating noise, shown in Fig.~\ref{rwalkers} (right). In three dimensions, the noise vector is rotated by a fixed angle $\beta$, along a randomly chosen great circle. The angle is related to the chosen correlation time $\Theta$ via the formula $\cos \beta = e^{-\Delta t/\Theta}$. The discrete soft noise function is thus
\begin{equation}\label{randforce}
\braket{\mathbf{W}(t)\cdot \mathbf{W}(t')} = \tanh\left(\frac{\Delta t}{2\Theta}\right)\frac{e^{-|t-t'|/\Theta}}{\Delta t}
\end{equation}
The normalization factor guarantees that for long separations $|t-t'| \gg \Theta$, the mean square displacement (MSD) is identical to the one of the Wiener process, thus satisfying the fluctuation-dissipation theorem. The time scale $\Theta$ introduced here is a high frequency cutoff necessary to be consistent with the short distance cutoff $\lambda$ of the Gaussian potential, Eq.~\ref{Gaussian}. The exact microscopic details of our model depend on the choice of $\Theta$ in addition to the potential Eq.~\eqref{Gaussian}. However, it is well known that long polymer chains exhibit universal behaviour on the large scale, and different molecular models produce the same results, up to a pre-factor determining the absolute units for time, distance, and molecular weight~\cite{glaser2014universality}. As a result, different models and chemical species can be compared by rescaling the units of measurement. However, when it comes to dynamical properties, the rescaling can only work if the chains do not unphysically cross each other over a timescale significantly longer than their longest relaxation time. In our model, the ratio of the excluded volume force to the soft random force scales as $\sqrt{\Theta/\tau}$ (see Eqs.~\eqref{randforce}, \eqref{Gaussian}), where $\tau = \lambda^2 \zeta/k_B T$ is the microscopic time unit. The frequency of chain crossings, hence the probability of overcoming this energy barrier, is expected to decrease as $\propto \exp\left(-\text{const.} \sqrt{\Theta/\tau}\right)$, which becomes negligible for big $\Theta$. The tradeoff is that the onset of entanglement $\tau_e$ is pushed to longer timescales, so it takes more steps to observe the same degree of topological confinement. The compromise that we have chosen here is $\Theta = 128\tau$ at which point we could not detect any chain crossings using direct geometrical analysis on a small system~\cite{korolkovas2016simulation}, as well as indirectly by observing chain dynamics on a big system. In the present study, the terminal relaxation time is seen to scale with $\tau_a \propto e^{N/N_a}$ for the very longest chains $N$ (see Fig.~\ref{msd}), in agreement with experiments, although no other multi-chain simulation could reach this regime so far. If our chains were crossable (the Rouse model), they would have relaxed many decades earlier, at a rate of $\tau_R \propto N^2$.

The excluded volume force can only be effective at suppressing chain crossings if the bond length $b$ between consecutive beads remains smaller than the repulsion range $\lambda$. Traditionally, this is imposed by a finitely extensible nonlinear elastic (FENE) spring force, but our soft time step $\Delta t = \tau$ is way too long for such a stiff force. Instead, we use a constraint algorithm similar to P-LINCS~\cite{hess2008p}. It maintains the bond distance at a chosen value of $b=0.7\lambda$, and although the solution is not exact, the residual fluctuations almost never exceed $b=\lambda$, which is the peak of the repulsive force guarding against chain crossings. Beyond that, there is a margin of error for up to $b=2\lambda$ when the repulsion falls to zero, but this has negligible probability and was never observed when testing the code.

The particle density in hardcore Lennard-Jones simulations is typically $0.85$. Our soft blobs on an approximate PIC grid must be set at a considerably lower density, and we have used $0.125$ blobs per pixel. This is to insure against fluctuations where too many chains come together and the force in the middle of such a cluster may cancel out, creating a possibility for an unphysical crossing. We have verified that we never have more than 3 blobs mapped to the same pixel, and even in such rare cases, typically 2 of them belong to the same chain.

The code of our simulation is written in CUDA and is available for download from Ref.~\cite{zenodoref}. The method is a useful addition to the polymer toolbox, thanks to its exceptional speed and the affordable price of GPU hardware. There is a possibility to extend this kind of softcore simulation to shear flow as well~\cite{korolkovas2018dynamical}, bearing in mind that the maximum shear rate $\kappa$ is limited to values much smaller than the inverse cutoff time: $\kappa \Theta \ll 1$.

\section{Results}
We have simulated a three-arm star polymer melt for a set of linearly increasing $N$: 112, 356, 600, 844, 1088, and 1332 beads per arm. The box size is fixed with a total of 110592 beads. The longest run was four months on a single Nvidia Titan XP.
\begin{figure}[tbh]
\includegraphics[width=\linewidth]{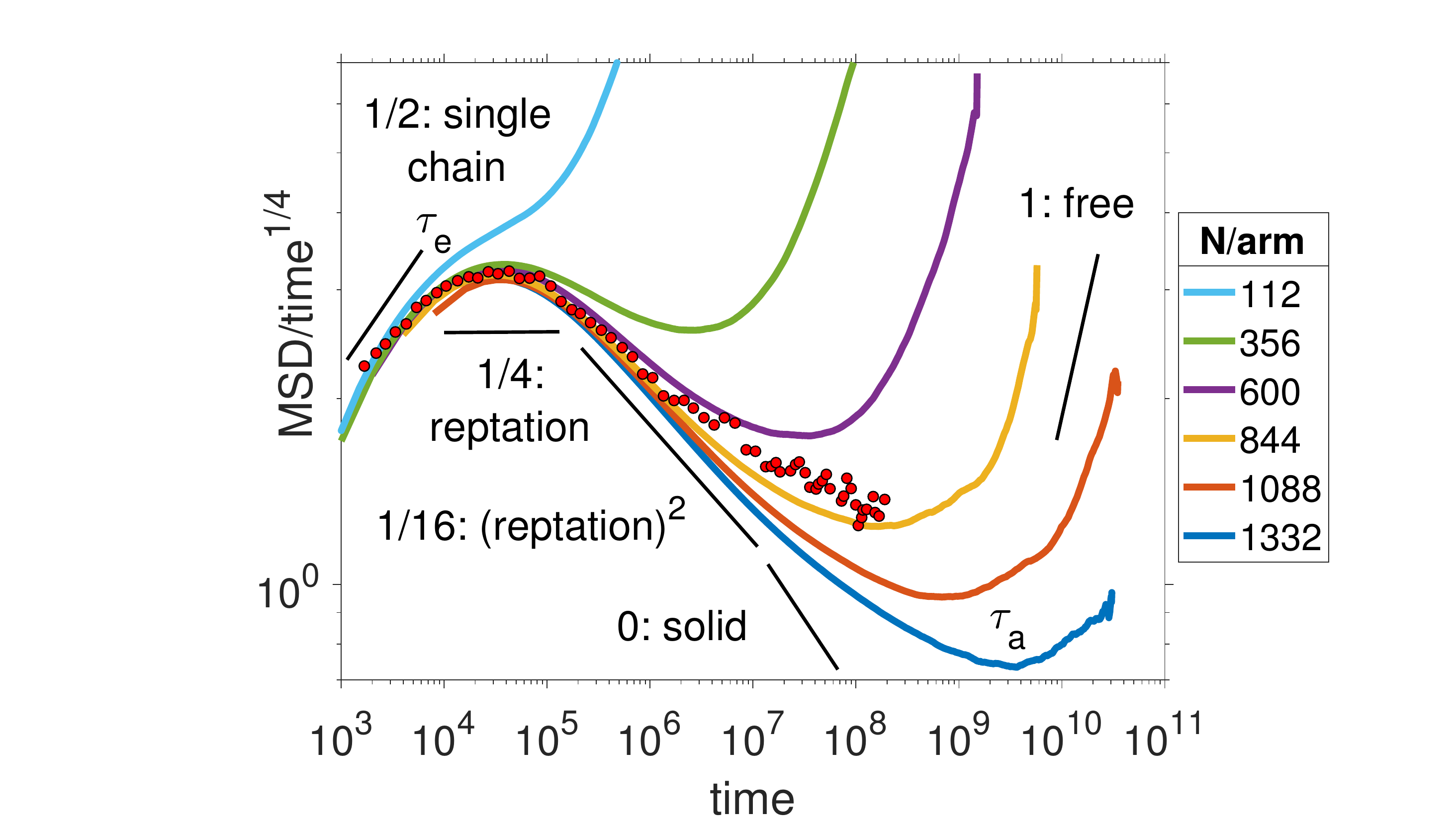}
\caption{Mean square displacement of the branch point, normalized by the reptation law $t^{1/4}$. The curves show our data for different arm lengths $N$, and the dots show the supercomputer result by Ba{\v{c}}ov{\'a} \emph{et al.}~\cite{bacova2013dynamics}. Solid black lines are theoretical predictions of terminal slopes.}\label{msd}
\end{figure}

The most topologically confined part of the molecule is the branch point. Its motion is quantified in terms of the mean square displacement (MSD):
\begin{equation}
g(t) = \braket{(\mathbf{R}(t)-\mathbf{R}(0))^2}
\end{equation}
shown in Fig.~\ref{msd}. The dataset spans over ten orders of magnitude in time and showcases a rich variety of dynamical phenomena. On the short time scale, all $N$ follow the same law of a free random walk $g \propto t/\tau$. On the very large time scale, it is again a random walk, but with a strong $N$ dependence of the pre-factor: $g \propto t/\tau_a$. For the three biggest $N$, we can fit the apparent arm retraction time to this equation: $\tau_a/\tau = \SI{1.33e6}{} e^{N/N_a}$, with $N_a = 170.7$. The exponential scaling is evident by the fact that the curves in Fig.~\ref{msd} become equidistant on the log scale, while the chain length $N$ is spaced out equally on a linear scale.

Connecting the micro- and the macroscopic worlds, is a very broad sub-diffusive regime. It is composed of smaller regions that follow well-defined scaling laws, marked by straight black lines in Fig.~\ref{msd}. The first is the Rouse model, describing a long linear isolated chain. It predicts the law of motion $g \propto (t/\tau)^{1/2}$ at the shallowest part~\cite{likhtman2012viscoelasticity}. When many uncrossable chains are brought together in a polymer melt, they become entangled and the MSD sinks down to $g \propto (t/\tau)^{1/4}$, which is the reptation law predicted by de Gennes~\cite{de1971reptation}.

Entangled star polymers have much slower dynamics than their linear counterparts. To clearly see the departure from reptation, the MSD data is divided by the $t^{1/4}$ law, so that a negative slope marks the onset of uniquely star dynamics. The central question is: what is the lowest possible slope in the limit of very large $N/N_a \gg 1$ stars? There is no firm consensus in the literature, while our data indicates a terminal value of $g\propto (t/\tau)^{1/16}$. It is the shallowest slope so far reported in entangled polymers.

The longest trajectory of star polymers available in the current literature is by Ba{\v{c}}ov{\'a} \emph{et al.}~\cite{bacova2013dynamics}, using the standard Kremer-Grest model, with an added bending stiffness to decrease the entanglement length. Their computation on the PRACE supercomputer has lasted 3-4 months (private communication). ESPResSo software package was used, which is a common choice for polymer simulations and has the performance similar to LAMMPS, another widely used option~\cite{grest2016communication}. The computational performance achieved by Ba{\v{c}}ov{\'a} \emph{et al.} is representative of the state-of-the-art capabilities across different polymer topics, like linear~\cite{hsu2017detailed} and ring polymers~\cite{halverson2012rheology}, which is roughly $10^9$ KG time steps per month.

We have represented their data of the branch point diffusion with red dots in Fig.~\ref{msd}. Because of widespread universality in polymer physics~\cite{zhang2015communication}, an apples-to-apples comparison is possible between the two models, provided we rescale the units of measurement. Our reference point is the first peak of the $g/t^{1/4}$ curve, signaling the departure from reptation (linear chain dynamics), to structurally slower inter-star dynamics. The time unit of  Ba{\v{c}}ov{\'a}'s data is multiplied by 17, and the distance unit by 3.55, matching up the peak rather well. Certainly, at short times $t\lesssim \Theta = 128\tau$ the two datasets do not agree because of our coarse-graining scheme, but at large $t\gg \Theta$ they converge to the same behaviour. Judging from the slope of the dots, their chain length falls at around $N \approx (600+844)/2=722$ in our units. In KG units, their length was reported at $N_{KG} = 200$, thus our soft chains have to be a factor of $722/200=3.6$ longer to have the same degree of topological confinement, and thereby the same large scale dynamics. Perhaps in the future this figure may be reduced if we implement bending stiffness as well. On the flip side, our soft time step is very long at $\Delta t = 1.0 \tau$, while the KG model requires $\Delta t = 0.01\tau_{KG}$ to resolve its stiff forces. Altogether, with an improved GPU algorithm, we can reproduce the same dynamics as Ref.~\cite{bacova2013dynamics} in roughly one day as opposed to $3+$ months.

\begin{figure}[tbh]
\includegraphics[width=\linewidth]{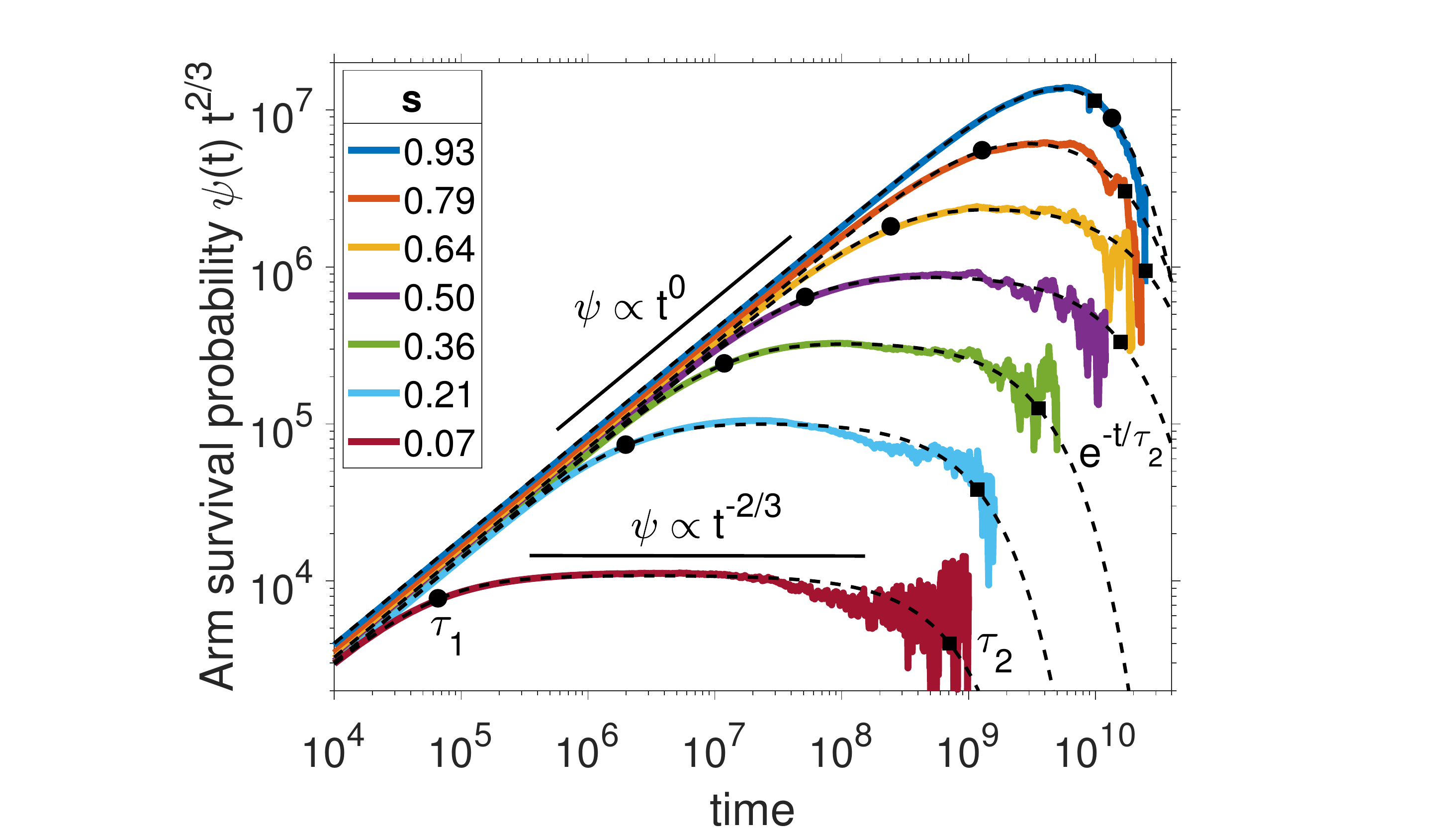}
\caption{The arm survival probability multiplied by $t^{2/3}$. Arm length $N=1088$. The monomer index ranges from $s=0.07$ close to the tip of the arm, to $s=0.93$ close to the branch point. Dashed lines show the fitting functions, Eq.~\eqref{fit}. Black dots mark the onset of relaxation $\tau_1(s)$, and black squares mark the end of the relaxation $\tau_2(s)$, values obtained from the fit.}\label{psi}
\end{figure}

This major improvement in the accessible time scale opens new avenues to the investigations of star polymer relaxation. The dynamics of the star as a whole is quantified by the arm survival probability $\psi(s,t)$. It is a function of the monomer index $s$, ranging from the tip of the arm, $s=0$, to the branch point, $s=1$. It counts the fraction of monomers that have not relaxed within a time interval $t$. This function is obtained by cross-correlating the local bond vector and the end-to-end vector of the arm:
\begin{equation}
\psi(s,t) = \braket{\partial \mathbf{R}(s,t)/\partial s\cdot [\mathbf{R}(1,0)-\mathbf{R}(0,0)]}
\end{equation}
Its overall relaxation spectrum $\tau(s)$ has been thoroughly discussed in the literature~\cite{milner1997parameter, mcleish2002tube, bacova2013dynamics}. Our dataset allows to extract more than just a single time scale. We report a universal law of $\psi \propto t^{-2/3}$, emerging for all values of $s$, valid for large $N$, across a broad range of time $\tau_1(s)\ll t \ll \tau_2(s)$. This finding is emphasized in Fig.~\ref{psi}, showing the rescaled function $\psi t^{2/3}$, so that the power law appears as a flat line. To fit the entire spectrum we propose this semi-empirical function:
\begin{equation}\label{fit}
\psi(s,t) = \frac{\psi_0 e^{-t/\tau_2}}{\left(1+(t/\tau_1)^{4/3} \right)^{1/2}}
\end{equation}
shown in dashed curves. The mixing exponent of $4/3\times 1/2 = 2/3$ was found to best represent the cross-over between $\psi \propto t^0$ and $\psi \propto t^{-2/3}$. An exponential tail $e^{-t/\tau_2}$ was added to model the final transition to zero, at $t\gg \tau_2(s)$. The combined Eq.~\ref{fit} is an excellent fit for all available data, including other chain sizes $N$ that are not shown here. Two characteristic timescales are needed: the onset $\tau_1(s)$ and the end $\tau_2(s)$ of relaxation, marked by black dots and squares, respectively. It appears that $\tau_2(s) \approx \tau_1(1)$ may actually be a constant, but more data is needed to confirm this speculation.

\section{Discussion}
The earliest theory of star polymer dynamics considered just a single star embedded in a network of fixed obstacles. Such a star could only relax by arm retraction, which was confirmed by computer simulation. However, this theory was far off the real world data of star melts, where the obstacles are other moving stars. To fix the discrepancy, dynamic tube dilation (DTD) was proposed. It is a hierarchical process, starting at the tip of the arm which is the most mobile and moves first, simultaneously relaxing an entanglement of a nearby star. The last to move is the branch point, and the probability of this happening is exponentially low, resulting in the terminal relaxation time $\tau_a \propto e^{N/N_a}$, but the pre-factor $N_a$ is smaller than for fixed obstacles. This theory~\cite{frischknecht2000self} reproduces large scale experimental observables~\cite{clarke2006self}, such as the overall viscosity~\cite{fetters1993rheological} and the self-diffusion coefficient~\cite{shull1990effect}. However, it does not fully agree with dielectric spectroscopy data which is more sensitive to the internal motions of the molecule~\cite{watanabe2000tube, watanabe2006constraint}. Further relaxation mechanisms like partial DTD and constraint release (CR) were introduced, showing promising results regarding linear rheology data~\cite{qiao2006constraint}.

However, regardless of the detailed mechanisms, all current theories maintain that the longest relaxation time is an exponential function of the molecular weight. As a consequence, the terminal diffusion slope (MSD over time) must asymptotically approach
\begin{equation}\label{slope}
\text{slope} \propto \frac{\log(R_g^2)}{\log(\tau_a)} \propto \frac{\log{N}}{N} \rightarrow 0
\end{equation}
for very large $N$, since the star radius of gyration follows a power law $R_g^2 \propto b^2 N$, similar to linear chains, and the relaxation occurs once $g(\tau_a) \approx R_g^2$. By contradiction, if the slope has a finite non-zero value, the predicted exponential growth of the relaxation time $\tau_a$ cannot last forever, and must eventually cross over to a power law. There is evidence for this in a recent simulation where the arm retraction mechanism has been artificially suppressed by immobilizing the arm tips~\cite{bacova2013dynamics}. Contrary to theory, the MSD of the branch point continued to grow, from which the existence of some unknown relaxation was inferred, perhaps deep diving nodes or end-looping constraint release.

The lowest MSD slope that we can confirm from our data is $t^{1/16}$. It is a plausible result for the true terminal value, although at this stage we cannot entirely exclude the possibility of an even lower slope. Computer simulations may be vulnerable to several biases: limited box size, insufficient equilibration time, unphysical chain crossings, and some technical issues like round-off errors that could start creeping up beyond $10^{10}$ steps. If despite our best efforts, any of these problems are present, the simulated dynamics will appear faster than expected.

Although experiments cannot directly measure it, the arm survival probability $\psi(s,t)$ is an important building block in polymer theories. They have mostly focused on the effective relaxation time spectrum $\tau(s)$, while the functional form of $\psi(s,t)$ itself has not been deliberated. To extract $\tau(s)$, previous simulations~\cite{bacova2013dynamics, holler2017role} have used phenomenological functions like the stretched exponential $\exp\left(-[t/\tau(s)]^\alpha\right)$. Their result corresponds to the start of the relaxation process, $\tau(s) \approx \tau_1(s)$. We are not aware of previous work that would have anticipated the ensuing broad power decay of $t^{-2/3}$, let alone the cross-over to an exponential tail at $\tau_2(s)$. At this stage we have no physical explanation for the observed $t^{-2/3}$ law. Further theoretical work is necessary to find a justification, and tie it into the MSD law of $t^{1/16}$.

\section{Acknowledgements}
The author thanks Anton Devishvili for valuable technical advice, Max Wolff, Franz A. Adlmann, Alexei Vorobiev and Philipp Gutfreund for scientific guidance. The Carl Tryggers stiftelse and the Swedish research council are acknowledged for financial support, grant CTS 16:519. The Titan XP used for this research was donated by the NVIDIA Corporation. 



\bibliography{manuscript}

\end{document}